\documentstyle[aps,prl,twocolumn,epsf]{revtex}
\begin{document}
\draft
\wideabs{
\title{Regimes of quantum degeneracy in trapped 1D gases}
\author{D.S. Petrov${}^{1,2}$, G.V. Shlyapnikov${}^{1,2}$, and J.T.M. Walraven${}^{1}$}
\address{${}^1$ FOM Institute for Atomic and Molecular Physics, Kruislaan 407, 
1098 SJ Amsterdam, The Netherlands \\
${}^2$ Russian Research Center, Kurchatov Institute, 
Kurchatov Square, 123182 Moscow, Russia}

\date{\today}
\maketitle
\begin{abstract}
We discuss the regimes of quantum degeneracy in a trapped 1D gas and 
obtain the diagram
of states. Three regimes have been identified:
the BEC regimes of a true condensate and quasicondensate, and the regime 
of a trapped gas of Tonks (gas of impenetrable bosons). The presence of a 
sharp cross-over to the BEC regime requires extremely small interaction 
between particles. We discuss how to distinguish between true and 
quasicondensates in phase coherence experiments.  
\end{abstract}
}
\narrowtext 

Low-temperature 1D Bose systems attract a great deal of interest
as they show a remarkable physics not encountered in 2D and 3D.
In particular, the 1D Bose gas with repulsive interparticle interaction
(the coupling constant $g>0$) becomes more non-ideal with decreasing  
1D density $n$ \cite{LiLi}.
The regime of a weakly interacting gas requires the correlation length
$l_c=\hbar/\sqrt{mng}$ ($m$ is the atom mass) to be much larger than the mean
interparticle separation $1/n$.
For small $n$ or large interaction, where this condition is violated, the
gas acquires Fermi properties as the wavefunction strongly decreases at short 
interparticle distances \cite{Girardeau,LiLi}. 
In this case it is called a gas of impenetrable bosons or gas of Tonks 
(cf. \cite{Tonks}).

Spatially homogeneous 1D Bose gases 
with repulsive interparticle interaction have been extensively studied in the
last decades. For the delta-functional interaction, 
Lieb and Liniger \cite{LiLi} have calculated the ground state energy
and the spectrum of elementary excitations which at low momenta turns out 
to be phonon-like. Generalizing the Lieb-Liniger approach, Yang and Yang
\cite{YY} have proved the analyticity of thermodynamical functions at any
finite  temperature,
which indicates the absence of a phase transition. However, 
at sufficiently low $T$ the correlation properties of a 1D Bose gas 
are qualitatively different from classical high-$T$ properties.
In the regime of a weakly interacting gas ($nl_c\gg 1$) the density 
fluctuations are suppressed \cite{KK}, whereas at finite $T$ the long-wave
fluctuations of the phase lead to exponential decay of
the one-particle density matrix at large distances \cite{KK,RC}. A
similar picture, with a power-law decay of the density matrix, was found
at $T=0$ \cite{Schwartz,Haldane}. Therefore, the Bose-Einstein condensate is
absent at any $T$, including $T=0$. Earlier studies of 1D Bose systems are
reviewed in \cite{Popov}. They allow us to conclude that in 1D  gases the
decrease of temperature leads to a continuous transformation of correlation
properties from ideal-gas classical to interaction/statistics dominated.
A 1D classical field model for calculating correlation
functions in the conditions, where both the density and phase fluctuations
are important, was developed in \cite{Sc} and with respect to Bose gases
in \cite{Castin}.
Interestingly,  1D gases can posses the property of superfluidity at $T=0$
\cite{Sonin,Popov}. Moreover, at finite $T$ one can have metastable 
supercurrent states which decay on a time scale independent of the size
of the system \cite{Kagan}. 

The earlier discussion of 1D Bose gases was mostly academic as there was no
possible realization of such a system. Fast progress in
evaporative and optical cooling of trapped atoms and the observation of 
Bose-Einstein 
condensation (BEC) in trapped clouds of alkali atoms \cite{discovery}
stimulated a search for non-trivial trapping geometries. At present, there are
significant efforts to create (quasi)1D trapped gases \cite{quasi1D}, where the
radial motion of atoms in a cylindrical trap is tightly confined and they only
undergo zero point radial oscillations. Then, kinematically the gas is 1D, and
the difference from purely 1D gases will only be related to the value of the
interparticle interaction which now depends on the radial confinement. 
The presence of the axial confinement allows one to
speak of a trapped 1D gas.

Ketterle and van Druten \cite{vDK} considered a trapped 1D ideal gas and have
revealed an essential role of the discrete structure of the trap levels. They have established that
at temperatures $T< N\hbar\omega/\ln{2N}$, where $N$ is the number of particles and $\omega$ the
trap frequency, the population of the ground state rapidly grows with decreasing $T$ and
becomes macroscopic. 
Thus, one has a clear BEC-like behavior of the ideal trapped 1D cloud.

A fundamental question concerns the influence of interparticle interaction
on the presence and nature of a Bose-condensed state and on the
character of a cross-over to the BEC regime.       
In this Letter we discuss the regimes of quantum degeneracy in a trapped 1D
gas with repulsive interparticle interaction. 
We find that the presence of a
sharp cross-over to the BEC regime, predicted in 
\cite{vDK}, requires extremely small interaction between particles. Otherwise,
the decrease of temperature leads to a continuous transformation of a classical
gas to quantum degenerate. We identify 3 regimes at $T\ll T_d$, where 
$T_d\approx N\hbar\omega$ is the degeneracy temperature. 
For a sufficiently large interparticle interaction and the number of particles
much smaller than a characteristic value $N_*$, at any $T\ll T_d$ one has 
a trapped gas of Tonks, with the density profile characteristic for an ideal 
Fermi gas. For $N\gg N_*$ we have a weakly interacting
gas.  
The presence of the trapping potential introduces a finite size of the sample
and drastically changes the picture of long-wave fluctuations of the phase
compared to the earlier discussed uniform case.
We calculate the density and phase fluctuations and find that well below $T_d$
there is a 
quasicondensate, i.e. the Bose-condensed state where 
the density fluctuations are suppressed but the phase still fluctuates. 
At very low $T$ the long-wave fluctuations of the phase are suppressed due to 
a finite size of the system, and we have a true condensate. The true
condensate and the quasicondensate have the same Thomas-Fermi density profile
and local correlation properties, and we analyze how to distinguish
between these Bose-condensed states in phase coherence experiments.

We first discuss the coupling constant $g$ for possible realizations of 
1D gases. These realizations
imply particles in a cylindrical trap, which are tightly confined in the 
radial ($\rho$) direction,
with the confinement frequency $\omega_0$ greatly exceeding the 
mean-field interaction. Then, at sufficiently low $T$ the radial motion 
of particles is
essentially ``frozen'' and is governed by the ground-state wavefunction of 
the radial harmonic oscillator. If the radial extension of the wavefunction,
$l_0=(\hbar/m\omega_0)^{1/2}\gg R_e$, where $R_e$ is the characteristic radius
of the interatomic potential, the interaction between particles 
acquires a 3D character and will be characterized by the 3D scattering length $a$. 
In this case, assuming $l_0\gg |a|$, we have
\begin{equation}      \label{cc}
g=2\hbar^2a/ml_0^2.
\end{equation}
This result follows from the analysis in \cite{Olshanii} and can also be
obtained by averaging the 3D interaction over the radial density profile. Thus,
statistical properties of the sample are the same as those of a purely 1D
system with the coupling constant  given by Eq.(\ref{cc}).  

In the regime of a weakly interacting gas, where $nl_c\gg 1$, we have a
small parameter  
\begin{equation} \label{small} 
\gamma=1/(nl_c)^2=mg/\hbar^2 n \ll 1. 
\end{equation} 
For particles trapped in a harmonic (axial) potential $V(z)=m\omega^2z^2/2$,
one can introduce a complementary dimensionless quantity
$\alpha=mgl/\hbar^2$ which provides a relation between the interaction strength 
$g$ and the trap frequency $\omega$ ($l=\sqrt{\hbar/m\omega}$ is the amplitude
of axial zero point oscillations). 

At $T=0$ one has a true condensate: In the Thomas-Fermi (TF) regime the mean square
fluctuations of the phase do not exceed $\sim \gamma^{1/2}$ and, hence, these 
fluctuations are small under the condition (\ref{small}) (see \cite{Ho}). The
condensate  wavefunction is determined by the Gross-Pitaevskii equation (GPE)
which gives the  well-known TF parabolic density profile $n_0(z)=n_{\rm
0m}(1-z^2/R_{TF}^2)$. The maximum  density $n_{\rm 0m}=n(0)=\mu/g$, the TF size
of the condensate $R_{TF}=(2\mu/m\omega^2)^{1/2}$, and the chemical potential
$\mu=\hbar\omega(3N\alpha/4\sqrt2)^{2/3}$.  For $\alpha\gg 1$ we are always in
the TF regime ($\mu\gg\hbar\omega$). In this case, Eq.(\ref{small}) requires a
sufficiently large number of particles: \begin{equation}    \label{TF}
N\gg N_*=\alpha^2.
\end{equation}
Note that under this condition the ratio $\mu/T_d\sim (\alpha^2/N)^{1/3}\ll 1$.
For $\alpha\ll 1$ the criterion (\ref{small}) of a weakly interacting gas is
satisfied at any $N$, and the condensate is in the TF regime if $N\gg\alpha^{-1}$.
In the opposite limit the mean-field interaction is much smaller than the level
spacing in the trap $\hbar\omega$. Hence, one has a macroscopic occupation of the ground
state of the trap, i.e. there is an ideal gas condensate with a Gaussian density profile. 

At this point, we briefly discuss the cross-over to the BEC regime, predicted by 
Ketterle and van Druten \cite{vDK}. They considered a trapped 1D ideal gas and found that
the decrease of temperature to below $T_c=N\hbar\omega/\ln{2N}$ leads to a
strong increase of the  population of the ground state, which rapidly becomes
macroscopic. This sharp cross-over originates from the discrete
structure of the trap levels and is not observed in quasiclassical calculations
\cite{Kleppner}. We argue that the presence of the interparticle  interaction
changes the picture drastically. One can only distinguish between the (lowest) 
trap levels if the interaction between particles occupying a particular level is
much smaller  than the level spacing. Otherwise the interparticle interaction
smears out the discrete  structure of the levels. For $T$ close to $T_c$ the
occupation of the ground state  is $\sim T_c/\hbar\omega\approx N/\ln{2N}$ \cite{vDK} and,
hence, the mean-field interaction  between the particles in this state (per
particle) will be $Ng/l\ln{2N}$. The sharp BEC cross-over 
requires this quantity to be much smaller than $\hbar\omega$, and we arrive at
the condition  $N/\ln{2N}\ll \alpha^{-1}$. For a realistic number of trapped
particles ($N\sim 10^3-10^4$)  this is practically equivalent to the condition
at which one has the ideal gas Gaussian  condensate (see above). 

As we see, the sharp BEC cross-over requires small $\alpha$. 
For possible realizations of 1D gases, using
the coupling constant $g$ (\ref{cc}), we obtain $\alpha=2al/l_0^2$. Then, even for the
ratio $l/l_0\sim 10$ and moderate radial confinement with $l_0\sim 1 \mu$m,
we have $\alpha\sim 0.1$ for Rb atoms ($a\approx 50$ \AA). Clearly, for a
reasonably large number of particles the cross-over condition $N\ll \alpha^{-1}$ can only
be fulfilled at extremely small interparticle interaction. One can think of reducing $a$ 
to below $1$\AA$\,$ and achieving $\alpha<10^{-3}$ by using Feshbach resonances as in the
MIT and JILA experiments \cite{Feshbach}. In this case, already for $N\sim 10^3$
one can expect the sharp BEC cross-over and the existence of the Gaussian
condensate at $T<T_c$.

We now turn to the case of large $\alpha$.
For $\gamma \gg 1$ one has a gas of Tonks
\cite{LiLi,com1}. The one-to-one mapping of this system to a gas of free
fermions \cite{Girardeau} ensures the fermionic spectrum and density
profile of a trapped gas of Tonks.  
For (axial) harmonic trapping the condition $\gamma\gg 1$ requires
$N\ll N_*$.
The chemical potential is equal to
$N\hbar\omega$, and the density distribution is 
$n(z)=n_m\sqrt{1-(z/R)^2}$, where $n_m=\sqrt{2N}/\pi l$, and the size of the
cloud $R=\sqrt{2N}l$. 
The density profile $n(z)$ is quite different from
both the profile of the zero-temperature condensate and the classical
distribution of particles, which provides a root for identifying the 
trapped gas of Tonks in future experiments. The interference effects 
and dynamical properties of this system are now a subject of theoretical
studies \cite{K,G}.
In Rb and Na this regime can be achieved for
$N\alt 10^3$ by the Feshbach increase of $a$ to $\sim 500$ \AA$\,$ and  using
$\omega\sim 1$Hz and optical radial confinement with $\omega_0\sim 10$kHz
($\alpha\sim 50$). 

Large $\alpha$ and $N$ satisfying Eq.(\ref{TF}), or small $\alpha$ and $N\gg \alpha^{-1}$,
seem most feasible in experiments with trapped 1D gases. 
In this case, at any
$T\ll T_d$ one has a weakly interacting gas in the TF regime.
Similarly to the uniform 1D case, the decrease of temperature to below $T_d$ 
continuously transforms a classical 1D gas to the regime of quantum degeneracy.
At $T=0$ 
this weakly interacting gas turns to the true TF condensate (see above).
It is then subtle to understand how the correlation properties change with temperature 
at $T\ll T_d$. For this purpose, we analyze the behavior 
of the one-particle density matrix by calculating the fluctuations of the density and phase.
We {\it a priori} assume small density fluctuations and prove this statement relying on the
zero-temperature equations for the mean density $n_0(z)$ and excitations.
The operator of the density fluctuations is (see, e.g. \cite{Shev}) 
\begin{equation} \label{operrho}
\hat n^{\prime}(z)=n_0^{1/2}(z)\sum_{j=1}^{\infty} i f_j^{-}(z) \hat{a}_j+h.c.,
\end{equation} 
where $ \hat{a}_j$ is the annihilation operator of the excitation
with quantum number $j$ and energy $\epsilon_j$, $f_j^{\pm}= u_j \pm v_j$, and the
$u,v$ functions of the excitations are determined by the same Bogolyubov-de Gennes equations 
as in the presence of the TF condensate.

The solution of these equations gives the spectrum $\epsilon_j=\hbar\omega\sqrt{j(j+1)/2}$
\cite{Ho,stringari} and the wavefunctions
\begin{equation} \label{fpm}
f_j^{\pm}(x)=\left( \frac{j+1/2}{R_{TF}}\right)^{1/2}\left[\frac{2\mu}{\epsilon_j}(1-x^2)
\right]^{\pm 1/2}P_j(x),
\end{equation}
where $j$ is a positive integer, $P_j$ are Legendre polynomials, and $x=z/R_{TF}$. 
For the mean square 
fluctuations of the density, $\langle(\delta\hat n(z,z'))^2\rangle=\langle(\hat n^{\prime}(z)-
\hat n^{\prime}(z'))^2\rangle$, we have
\begin{eqnarray}     
\frac{\langle(\delta\hat n(z,z'))^2\rangle}{n_{0m}^2}=\sum_{j=1}^{\infty}\frac{\epsilon_j
(j+1/2)}{2\mu n_{0m}R_{TF}}(P_j(x)-P_j(x'))^2(1+2N_j), \nonumber 
\end{eqnarray}   
with $N_j=[\exp(\epsilon_j/T)-1]^{-1}$ being the occupation numbers for the excitations. At $T\gg\hbar\omega$
the main contribution to the density fluctuations comes from quasiclassical excitations ($j\gg 1$). 
The vacuum fluctuations are small: $\langle(\delta\hat n(z,z'))^2\rangle_0\sim n_{0m}^2\gamma^{1/2}$.
For the thermal fluctuations on a distance scale $|z-z'|\gg l_c$, we obtain
\begin{equation} \label{dfl1} 
\frac{\langle (\delta n(z,z'))^2\rangle_T}{n_{0m}^2}\approx
\frac{T}{T_d}\min \left\{\frac{T}{\mu},1\right\}.
\end{equation}

We see that the density fluctuations are strongly suppressed at temperatures $T\ll T_d$.
Then, one can write the total field operator as $\hat\psi(z)=\sqrt{n_0(z)}\exp(i\hat\phi(z))$,
where $\hat\phi(z)$ is the operator of the phase fluctuations, and the one-particle density matrix 
takes the form (see, e.g. \cite{Popov})
\begin{equation}  \label{opdm}
\!\langle\hat\psi^{\dagger}(z)\hat\psi(z')\rangle\!=\!\sqrt{n_0(z)n_0(z')}
\exp\{-\langle [\delta\hat\phi(z,z')]^2\rangle/2\},\!  \nonumber
\end{equation}
with $\delta\hat\phi(z,z')=\hat\phi(z)-\hat\phi(z')$. The operator of the phase fluctuations 
is given by (see \cite{Shev})
\begin{equation}      \label{operphi}    
\hat\phi(z)=[4n_0(z)]^{-1/2}\sum_{j=1}^{\infty}f_j^{+}(z)\hat a_j +h.c.,
\end{equation}
and for the mean square fluctuations we have
\begin{eqnarray}    
\langle[\delta\hat\phi(z,z')]^2\rangle=\sum_{j=1}^{\infty}\frac{\mu (j+1/2)}{2\epsilon_j n_{0m}R_{TF}}
(P_j(x)-P_j(x'))^2(1+2N_j).   \nonumber
\end{eqnarray}
For the vacuum fluctuations we find (c.f. \cite{Lutt,Ho})
$$\langle[\delta\hat\phi(z,z')]^2\rangle_0\approx (\gamma^{1/2}/\pi)\ln{(|z-z'|/l_c)},$$
and they are small for any realistic size of the gas cloud. 
The thermal fluctuations of the phase are mostly provided by the contribution of the lowest excitations.
A direct calculation, with $N_j=T/\epsilon_j$, yields
\begin{equation} \label{phi}
\langle[\delta\hat\phi(z,z')]^2\rangle_T=\frac{4T\mu}{3T_d\hbar\omega}\left|\log
\left[\frac{(1-x')}{(1+x')}\frac{(1+x)}{(1-x)}\right]\right|.
\end{equation}
In the inner part of the gas sample the logarithm in Eq.(\ref{phi}) is
of order unity.

Thus, we can introduce a characteristic temperature
\begin{equation} \label{Tph}
T_{ph}=T_d\hbar\omega/\mu
\end{equation}   
at which the quantity $\langle[\delta\hat\phi(z,z')]^2\rangle\approx 1$ on a distance scale
$|z-z'|\sim R_{TF}$. The characteristic radius of phase fluctuations is $R_{\phi}
\approx R_{TF}(T_{ph}/T)\propto N^{2/3}/T$, and for $T<T_{ph}$ it exceeds the sample size
$R_{TF}$.  This means that at $T\ll T_{ph}$ both the density and phase
fluctuations are suppressed, and  there is a true condensate. The
condition (\ref{TF}) always provides the ratio $T_{ph}/\hbar\omega\approx
(4N/\alpha^2)^{1/3}\gg 1$. 

In the temperature range, where $T_d\gg T\gg T_{ph}$, the density fluctuations are suppressed, but 
the phase fluctuates on a distance scale $R_{\phi}\ll R_{TF}$.
Thus, similarly to the quasi2D case \cite{our2d}, we have a condensate with
fluctuating phase (quasicondensate). The radius of the phase fluctuations
greatly exceeds the correlation length: $R_{\phi}\approx l_c(T_d/T)\gg l_c$.
Hence, the quasicondensate has the same
density profile as the true condensate. Correlation properties at distances
smaller than $R_{\phi}$ are also the  same. However, the phase fluctuations
lead to a drastic difference in the phase coherence properties. 

In Fig.1. we present the state diagram of the trapped 1D gas for $\alpha=10$ ($N_*=100$). 
For $N\gg N_*$, the decrease of temperature to below $T_d$ leads to 
the appearance of a quasicondensate which at $T<T_{ph}$ turns to the true condensate. In the $T-N$
plane the approximate border line between the two BEC regimes is determined by the equation
$(T/\hbar\omega)=(32N/9N_*)^{1/3}$. For $N<N_*$ the system can be 
regarded as a trapped gas of Tonks.

\begin{figure} 
\epsfxsize=\hsize 
\epsfbox{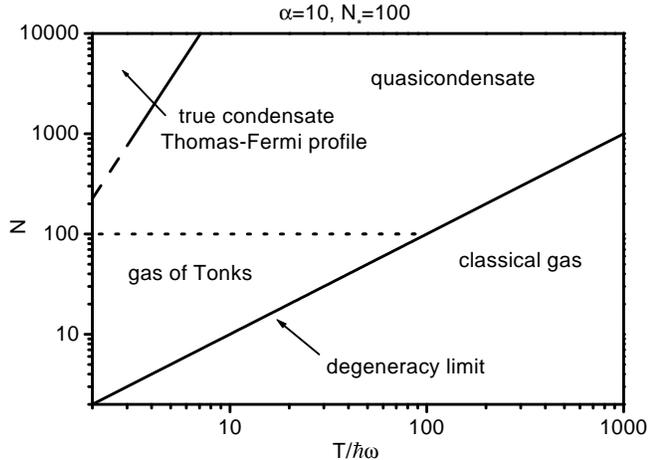} 
\caption{\protect
Diagram of states for a trapped 1D gas.   } 
\label{1} 
\end{figure} 

Phase coherence properties of trapped 1D gases can be studied in `juggling' 
experiments
similar to those with 3D condensates at NIST and Munich 
\cite{Phi,Ess}. Small clouds of atoms are ejected from the main cloud by stimulated Raman
or RF transitions. Observing the interference between two clouds, simultaneously ejected from
different parts of the sample, allows  the reconstruction of the spatial phase correlation
properties. Similarly, temporal correlations of the phase can be studied by overlapping clouds
ejected at different times from the same part of the sample. In this way juggling experiments
provide a direct measurement of the one-particle density matrix. Repeatedly juggling clouds of a
small volume $\Omega$ from points $z$ and $z'$ of the sample, for equal time of flight to the
detector we have the averaged detection signal 
$I=\Omega[n_0(z)+n_0(z')+2\langle\hat\psi^{\dagger}(z)\psi(z')\rangle]$.  

At $T\ll T_{ph}$ the phase fluctuations are small and one has a true
condensate. In this case, for $z'=-z$ we have
$\langle\hat\psi^{\dagger}(z)\hat\psi(z')\rangle=n_0(z)$ and
$I=4\Omega n_0(z)$, and there is a pronounced
interference effect: The detected signal is
twice as large as the number of atoms in the ejected clouds. 
The phase fluctuations grow with $T$ and for $T>T_{ph}$, where the true
condensate turns to a quasicondensate, the detection signal decreases as
described by $\langle\hat\psi^{\dagger}(z)\hat\psi(z')\rangle$
from Eqs.~(\ref{opdm}) and (\ref{phi}).  For $T\gg T_{ph}$ the phase
fluctuations completely destroy the interference between the two ejected
clouds, and $I=2\Omega n_0(z)$. 

In conclusion, we have identified 3 regimes of quantum degeneracy in a trapped
1D gas: the BEC regimes of a quasicondensate and true condensate, and the 
regime of a trapped gas of Tonks. The creation 
of 1D gases will open handles on interesting phase coherence studies and
the studies of ``fermionization'' in Bose systems.

This work was financially supported by the Stichting voor Fundamenteel Onderzoek der Materie (FOM), by INTAS, and by 
the Russian Foundation for Basic Studies.


\begin{references}
\bibitem{com1} The realization of the axially uniform gas of Tonks in
a square-well (axial) potential was discussed in \cite{Olshanii}.
\end{references}
\end{document}